\documentclass{rspublic}
\usepackage{epsfig, amssymb, amsmath, multicol}

\newcommand{\ea}{{\it et al. }}
\newcommand{\apj}{{\it Astrophys. J.}}
\newcommand{\apjs}{{\it Astrophys. J. Suppl. Ser.}}
\newcommand{\aj}{{\it Astron. J.}}
\newcommand{\mnras}{{\it Mon. Not. R. Astron. Soc.}}
\newcommand{\aanda}{{\it Astron. Astrophys.}}
\newcommand{\pasp}{{\it Publ. Astron. Soc. Pac.}}
\newcommand{\ptra}{{\it Phil. Trans. R. Soc. A}}

\begin{document}

\title[Star clusters as laboratories for stellar and dynamical evolution]{Star clusters as laboratories for \\ stellar and dynamical evolution}

\author[J. S. Kalirai \& H. B. Richer]{Jason S.\ Kalirai$^1$ and Harvey B.\ Richer$^2$}

\affiliation{$^1$Space Telescope Science Institute, 3700 San Martin Drive, Baltimore MD 21218, USA \\
$^2$Department of Physics and Astronomy, University of British Columbia, 6224 Agricultural Road, 
Vancouver BC V6T~1Z1, Canada}

\label{firstpage}

\maketitle


\begin{abstract}{Star clusters, stellar evolution, white dwarfs}
Open and globular star clusters have served as benchmarks for the
study of stellar evolution due to their supposed nature as simple
stellar populations of the same age and metallicity. After a brief
review of some of the pioneering work that established the importance
of imaging stars in these systems, we focus on several recent studies
that have challenged our fundamental picture of star clusters.  These
new studies indicate that star clusters can very well harbour multiple
stellar populations, possibly formed through self-enrichment processes
from the first-generation stars that evolved through
post-main-sequence evolutionary phases.  Correctly interpreting
stellar evolution in such systems is tied to our understanding of both
chemical-enrichment mechanisms, including stellar mass loss along the
giant branches, and the dynamical state of the cluster.  We illustrate
recent imaging, spectroscopic and theoretical studies that have begun
to shed new light on the evolutionary processes that occur within star
clusters.
\end{abstract}

\section{An historic look at stellar evolution and star clusters}

\begin{figure}[h]
\begin{center}
\includegraphics[width=7.4cm, bb= 20 145 575 575]{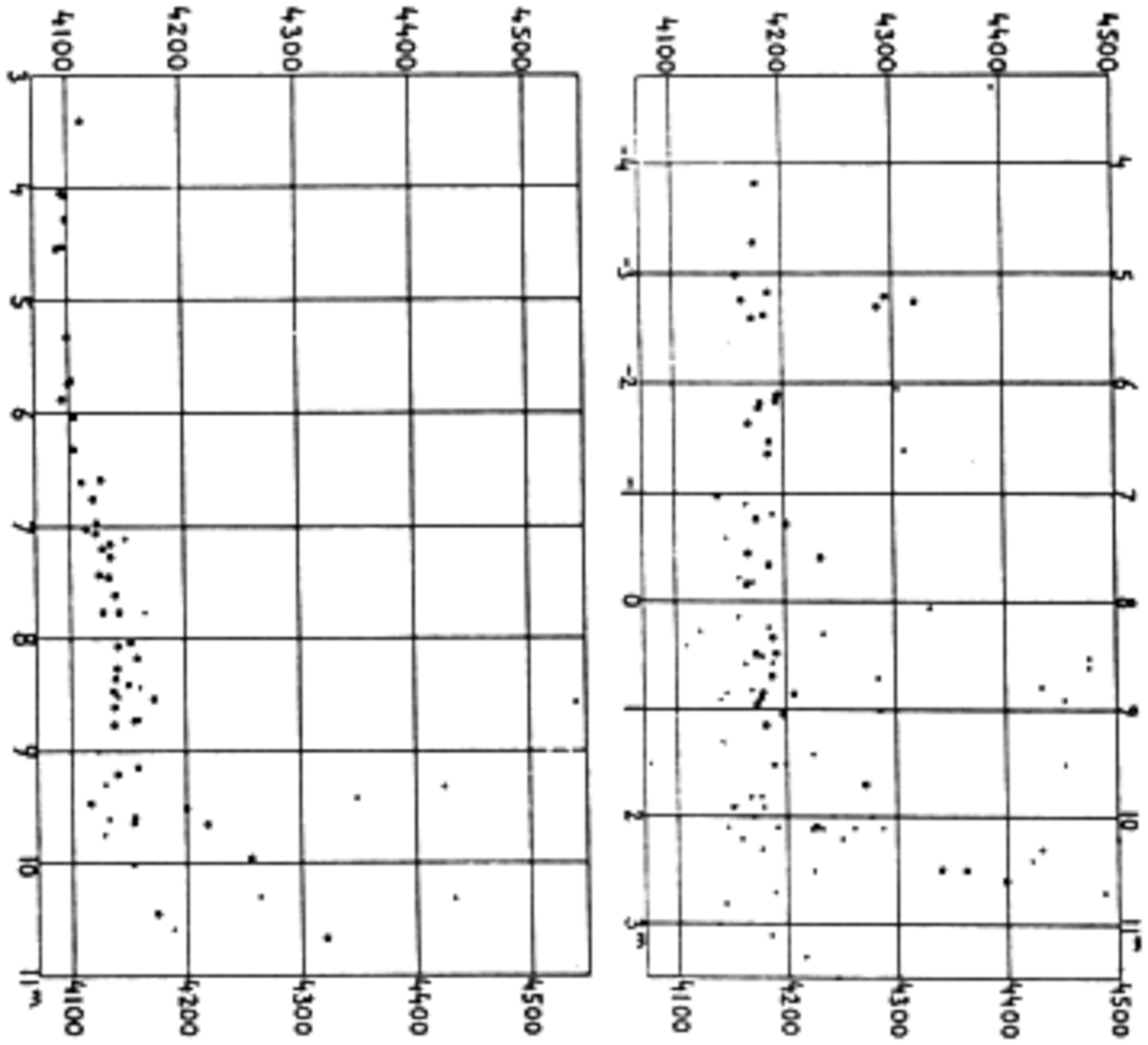}
\includegraphics[width=5.1cm]{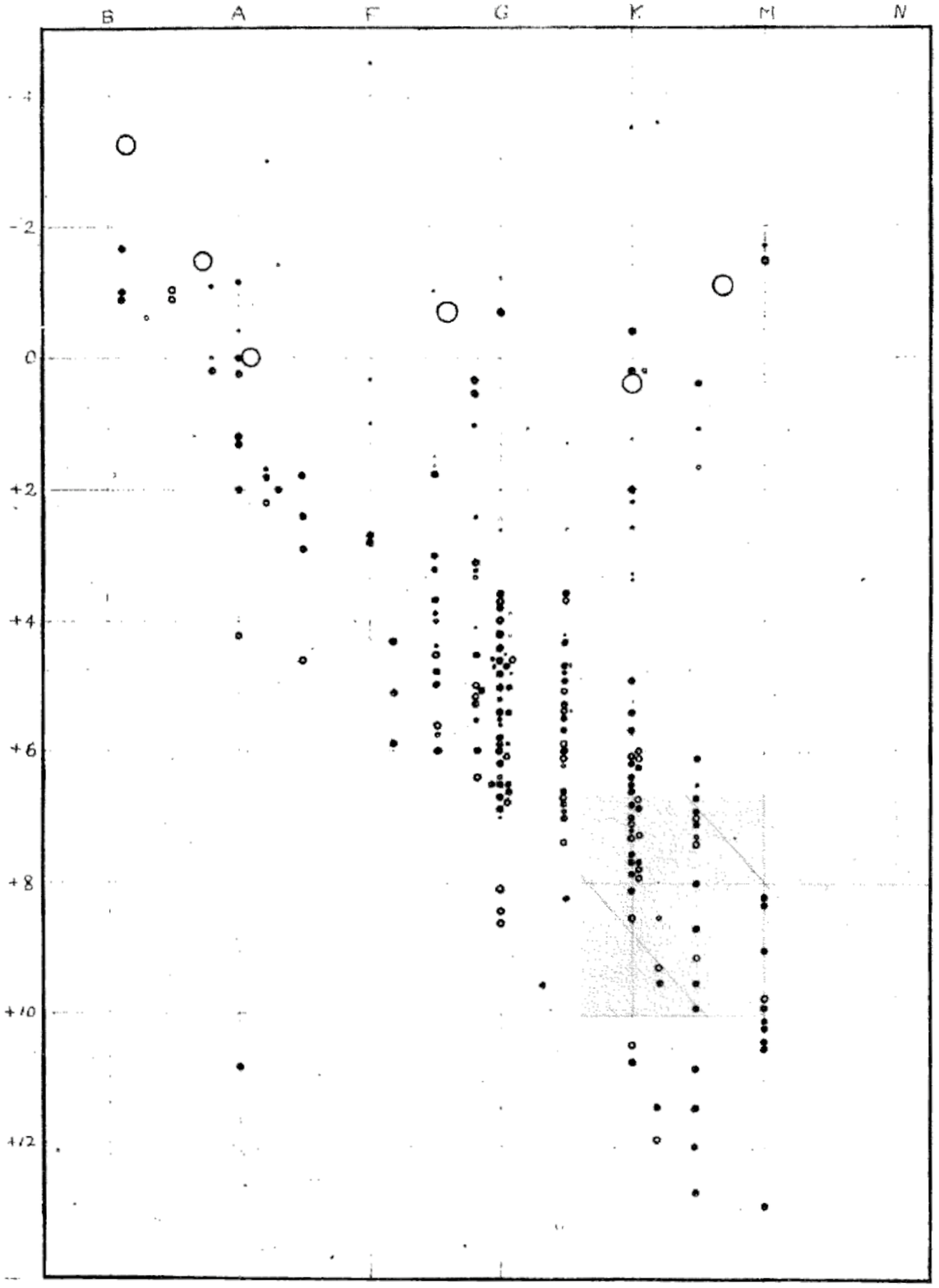}
\end{center}
\caption{The first published Hertzsprung--Russell (H--R) diagrams of
the Pleiades and Hyades clusters ({\it left} and {\it middle};
Hertzsprung 1911), as well as for nearby field stars ({\it right};
Russell 1914{\it a}).  Plotted on the vertical axis are the magnitudes
(e.g., more luminous stars are at the top) and on the horizontal axis
are the peak wavelengths or spectral types of the stars (e.g., hotter,
bluer stars towards the left).\label{fig:HRDiagrams}}
\end{figure}

Understanding stellar evolution has been one of the most important
pursuits of observational astronomy over the past century.  In 1905,
Ejnar Hertzsprung demonstrated that stars emitting light with a
similar spectrum, and with the same parallax, could have very
different luminosities.  Hertzsprung referred to these nearby stars,
which therefore had the same temperature and distance, as `giants'
(high luminosity) and `dwarfs'.  Just a few years after this study,
Hertzsprung and Henry Norris Russell obtained and analysed new
observations that characterized the basic properties of hundreds of
nearby field stars (Russell 1913; 1914{\it a}) and for the nearest
co-moving groups (e.g., the Pleiades and Hyades: see
figure~\ref{fig:HRDiagrams}; Hertzsprung 1911).  This pioneering work
led to one of the most important diagnostic tools in astrophysics, the
luminosity versus spectral type plane, rightfully named the
Hertzsprung--Russell (H--R) diagram.

The first H--R diagrams illustrated clear evidence of groupings and
sequences of stars, leading to speculation of the probable order of
stellar evolution, as well as some interesting correlations (Russell
1914{\it a,b}).  For example, based on these observations, Russell
remarks that ``there appears, from the rather scanty evidence at
present available, to be some correlation between mass and
luminosity''.  Remarkably, these initial observations were rather
complete, illustrating a rich main sequence extending over 10
magnitudes from A-type stars to M dwarfs (including several eclipsing
variables), an abundant population of red giants, and the first white
dwarf observed (40~Eri~B, seen by W.\ Herschel in
1783).\footnote{Given the lack of other `faint white stars', Russell
initially questioned the quality of the spectrum of this star, which
is a member of a triple system (Russell 1914{\it a}).}

Star clusters were recognized as excellent laboratories to test
theories of stellar evolution soon after the first H--R diagrams of
the Pleiades and Hyades were published.  Over the years, it was
quickly realized that each individual cluster is a simple stellar
population, being coeval, cospatial and isometallic (some of this
knowledge came later) and, therefore, represents a controlled testbed
with well-established properties that are common to all member stars
(see also Bruzual 2010).  Yet, the constituent stars in any given
cluster span a large range in stellar mass, the most important
parameter dictating stellar evolution as we now know.  The present
observations of an H--R diagram for any star cluster therefore
represents a snapshot of how stellar evolution has shaped the
population at a given age and metallicity; the other properties only
have a secondary influence. As we link together the observations of
many of these clusters, over a wide range in age and metallicity, a
complete picture of stellar evolution begins to emerge.

Fortunately, the timing of Hertzsprung and Russell's first studies of
the H--R diagrams of nearby star clusters coincided with the
construction of the first large reflecting telescopes.  The Mount
Wilson Observatory's 1.5~m (1908) and 2.5~m (1917) facilities and the
Dominion Astrophysical Observatory's 1.8~m (1918) offered the exciting
possibility of undertaking the first systematic (resolved) studies of
many of the clusters in Messier's famous catalog (Messier
1774).\footnote{Available online at {\tt
http://messier.obspm.fr/xtra/history/m-cat71.html}.}  In a series of
papers, Harlow Shapley catalogued photometry for hundreds of
individual stars in the nearest star clusters (e.g., Shapley 1917; and
related articles in the 1910s and 1920s) and provided the first clues
as to the properties of these systems.  For example, Shapley measured
the distances and distributions in space, sizes, luminosities and even
proper motions of several nearby star clusters.  These photometric
studies led directly to the first ideas on stellar physics (e.g., the
static main sequence and the mass--luminosity relation; Eddington
1926), well before the source of a star's energy was known.  It was
only later, in the 1930s and 1940s, that hydrogen fusion and the
proton--proton chain were detailed, leading to much of our current
picture of stellar evolution (e.g., Chandrasekhar 1939; Bethe \&
Marshak 1939; Hayashi 1949; Henyey \ea 1959).

\section{Present-day H--R diagrams of open and \\ globular star clusters}

Star clusters are now understood to be the end products of
star-formation processes such as those we see in regions like the
Orion OB association (e.g., Warren \& Hesser 1977).  The clusters form
from the collapse and fragmentation of a turbulent molecular cloud
(e.g., Harris \& Pudritz 1994; Elmegreen \ea 2000; Bate \ea 2003; see
also Clarke 2010; Lada 2010).  Modern-day observations of Milky Way
star clusters generally group these systems into two classes, the
Population I open clusters that have lower total mass (tens to
thousands of stars) and are mostly confined to the Galactic disc
(Janes \& Adler 1982), and Population II globular clusters (tens of
thousands to hundreds of thousands or more stars) that are much more
massive and make frequent excursions into the Galactic halo (Zinn \&
West 1984; Harris 1996).

\begin{figure}[t]
\begin{center}
\includegraphics[width=9.2cm, angle=270]{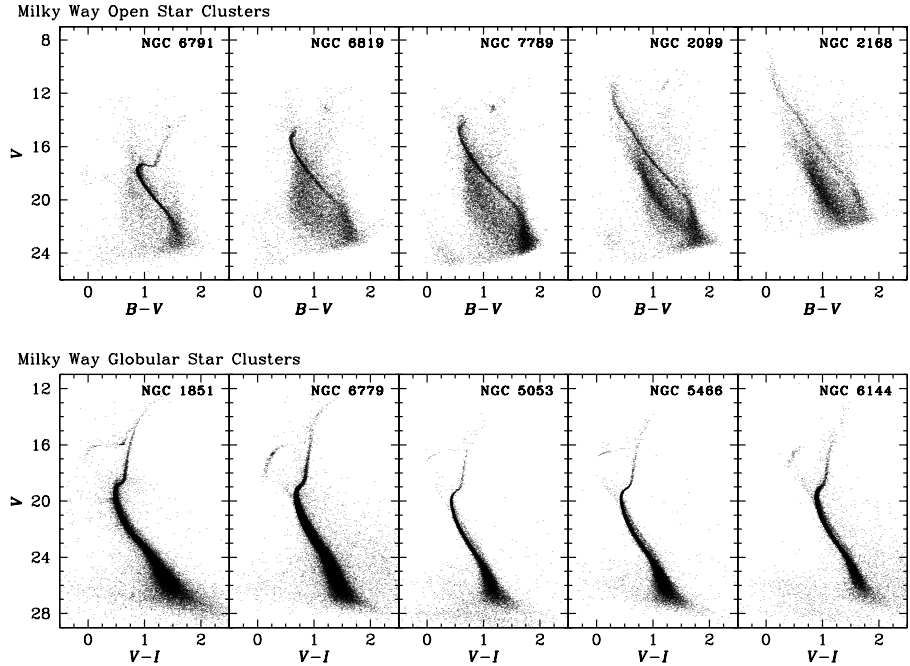}
\end{center}
\caption{{\it (top)} H--R diagrams of five rich open star clusters
observed as a part of the {\sl Canada--France--Hawaii Telescope} Open
Star Cluster Survey (Kalirai \ea 2001{\it a}).  The clusters are
arranged from the oldest in the top left-hand corner (8~Gyr) to the
youngest in the top right-hand corner (200~Myr).  {\it (bottom)} H--R
diagrams of five relatively unexplored globular clusters from the ACS
Survey of Galactic Globular Clusters (Sarajedini \ea 2007).  Note that
the photometry for some of the brightest red giants in these H--R
diagrams suffers from saturation effects. \label{fig:8cmds2}}
\end{figure}

Despite their importance for several aspects of stellar astrophysics,
including an understanding of stellar evolution, many rich Milky Way
star clusters have been historically neglected from observational
studies given their large angular sizes, great distances and/or heavy
extinction.  For those systems that have been studied, the existing
photometry often only extends to $V \sim 20$ mag, limiting scientific
studies to just the giants and brighter main-sequence stars of the
cluster.  The advent of wide-field CCD cameras on 4m-class telescopes
has recently provided us with a wealth of new data on the open
clusters of our Galaxy.  For example, projects such as the {\sl WIYN}
(Mathieu~2000) and {\sl CFHT} (Kalirai \ea 2001{\it a}) Open Star
Cluster Surveys have systematically imaged nearby northern-hemisphere
clusters in multiple filters down to a depth of $V = 25$ mag, enabling
a number of new studies.\footnote{A database of Galactic open cluster
H--R diagrams is available at\newline {\tt
http://www.univie.ac.at/webda}; Mermilliod~(1995).}  Similarly, the
Advanced Camera For Surveys (ACS) Survey of Galactic Globular Clusters
has provided homogeneous imaging of a large fraction of the Milky
Way's globulars that lacked previous high-resolution observations
(Sarajedini \ea 2007).\footnote{A database of Galactic globular
clusters is available at\newline {\tt
http://venus.mporzio.astro.it/$\sim$marco/gc/}.}  We illustrate the
quality of these new data in figure~\ref{fig:8cmds2} for a sample of
five very rich open star clusters (Kalirai \ea 2001{\it b,c}, 2003,
2007{\it a}, 2008) and five poorly studied globular clusters
(Sarajedini \ea 2007).

\subsection{Inverting H--R diagrams to constrain stellar evolution theory}

Interpreting stellar evolutionary processes from observations of star
clusters ideally involves comparing the observed H--R diagrams of
these systems to synthetic diagrams produced from theoretical stellar
evolution models (e.g., Aparicio \ea 1990; Tosi \ea 1991; Skillman \&
Gallart 2002).  The synthetic H--R diagram is constructed based on
Monte Carlo extractions of (mass, age) pairs, given a stellar initial
mass function, star-formation law and time interval for the
star-formation activity.  Each extracted synthetic star is placed in
the observed H--R diagram by interpolation of the adopted stellar
evolution tracks, and is given a photometric error based on the actual
measured errors of stars at that brightness.  Of course, any star is
only retained in the analysis if it actually would have been detected
in the observations, as calculated from incompleteness corrections
based on artificial-star tests.

For a given set of models, the comparison above can yield tight
constraints on the fundamental properties of the star cluster, such as
its age, metallicity, binary fraction and mass.  By extending these
comparisons to multiple sets of models, e.g., the Padova (Girardi \ea
2002), Yale--Yonsei (Demarque \ea 2004), VandenBerg (VandenBerg \ea
2006) or Dartmouth (Dotter \ea 2008) groups, the different assumptions
that are invoked to explain processes that do not come from first
principles can be tested.  For example, depending on the treatment of
core rotation, diffusion, gravitational settling and core
overshooting, the location of points in the synthetic H--R diagram
will vary and the distribution of points along various evolutionary
stages will be unique (see Kalirai \& Tosi 2004 for some comparisons).

\subsection{The present study}

There exist excellent reviews on the constraints that evolutionary
sequences in H--R diagrams have provided for the theory of how stars
evolve.  For example, Renzini \& Fusi~Pecci~(1988) highlighted our
understanding 20 years ago of the hydrogen-burning main sequence (both
for low- and higher-mass stars), subgiant branch, red-giant branch,
horizontal branch and asymptotic giant branch, and subsequent
evolution on to the white-dwarf cooling sequence (see also Iben 1965,
1967).

In this article, we shift focus to new inferences on stellar evolution
that have emerged from some of the most accurate and deepest H--R
diagrams constructed to date ({\it of any stellar population}).  For
example, very accurate photometric observations of several globular
clusters have now revealed that these stellar systems may not be the
simple populations that we have assumed for so many years and, in
fact, may contain sets of stars with different initial conditions.
Ultradeep imaging of the nearest systems has, for the first time,
provided us with a complete inventory of the stellar species in these
clusters, extending from the hydrogen-burning limit, through the
turnoff, to the brightest giants and down to the faintest cluster
remnants.  By combining these imaging data with new constraints from
spectroscopy, links have been made to directly connect the properties
of the main-sequence stars to their eventual white-dwarf state, and
therefore to probe stellar mass-loss rates over a wide range of
environments.  We also discuss the importance of coupling our basic
knowledge of stellar evolution, as gauged from observations of star
clusters, to the dynamical evolution of these clusters.  This includes
a look at how the death of stars in these clusters can affect the
global properties of the system and the evolution of stars within.
Finally, we draw some links on how the knowledge that we gain from
these local calibrators can directly influence the interpretation of
photometry of galaxies in the distant Universe.  We conclude with an
encouraging discussion of the exciting possibilities for advancing our
understanding of stellar evolution from new surveys that are
envisioned in the coming decade.

\section{Multiple stellar populations in individual star clusters} \label{multiple}

For a coeval system, the present-day stars on the main sequence and in
post-main-sequence evolutionary phases may be affected by the
first-generation evolving stars.  For example, the ejecta from the
evolution of the most massive stars in the cluster, i.e., the first
supernovae, may pollute the atmospheres of neighbouring stars.  As we
will see in \S\,\ref{massloss}, even stars only a few times more
massive than the Sun will lose 75\% of their mass in just a few
hundred million years after they evolve off the main sequence.  This
expelled material, e.g., from the asymptotic giant branch, was subject
to nucleosynthesis at the base of the convective envelope and can
therefore act as another source of pollution for less-evolved stars in
a cluster environment.  Indeed, evidence of variations in chemical
species among the red giants of nearby globular clusters has existed
for over 30 years (e.g., Cohen 1978; Peterson 1980; Norris \ea 1981;
Gratton 1982).  More recently, high-resolution spectroscopic studies
have been extended to unevolved stars with similar results, as
summarized in excellent review papers by Kraft (1994) and Gratton \ea
(2004).  This work suggests that self-enrichment of the intracluster
medium and of unevolved stars in globular clusters can produce light
elements (e.g., C, N, Na, O, Mg and Al) that are altered in their
ratios due to proton-capture reactions, a scenario that is not seen
among field halo stars.

The spectroscopic work on chemical abundances discussed above hints
that globular clusters may not be as simple and clean as we naturally
assume.  A remarkable observation by D'Antona \ea (2005) changed the
paradigm completely.  By using the Wide Field Planetary Camera 2
(WFPC2) aboard the {\sl Hubble Space Telescope (HST)}, D'Antona \ea
measured the thickness of the main sequence of the massive cluster
NGC~2808 and, for the first time, found photometric evidence of
multiple stellar populations on the main sequence.  Using techniques
similar to those described above, D'Antona \ea carefully characterized
their photometric errors on the main sequence and found that 20\% of
the cluster stars are bluer than expected from synthetic H--R
diagrams.  They explained the results by invoking a scenario where the
helium abundance of these stars was much larger than for the bulk of
the main-sequence stars.  D'Antona \ea go on to suggest {\it three}
phases of star formation in the cluster, with different helium
enhancement in each phase, from the winds of previous-generation,
massive, asymptotic-giant-branch stars (note that the helium yield is
a robust prediction from asymptotic-giant-branch evolutionary models;
Ventura \ea 2001).  Interestingly, enhanced helium abundances will
also lead to bluer tails in a cluster's horizontal-branch morphology,
suggesting a possible connection between the main-sequence
observations and the already observed dichotomy of NGC~2808's
horizontal branch (see also D'Antona \& Caloi 2004).  Of course, as
has been outlined extensively in the literature, the morphology of a
cluster's horizontal branch may also (or entirely) depend on several
other factors such as age, metallicity, mass loss, late flashers and
rotation (see Sweigart \& Catelan 1998).

\begin{figure}[h]
\begin{center}
\leavevmode 
\includegraphics[width=8.0cm]{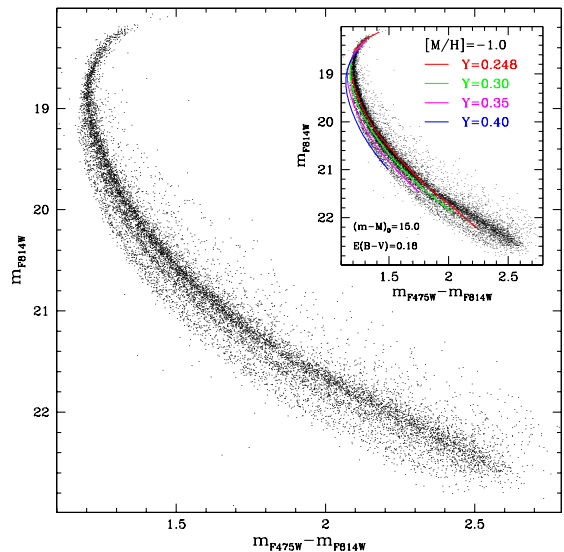}
\end{center}
\caption{The triple main sequence of the globular cluster NGC~2808,
taken from Piotto \ea (2007).  The stars are all proper-motion members
of the cluster and a correction has been made for differential
reddening along the line of sight.  {\it (inset)} Theoretical
isochrones for an age of 12.5~Gyr, with different helium content.
\label{fig:NGC2808}}
\end{figure}

Verification of the picture above came from extremely precise {\sl
HST}/ACS imaging observations of NGC~2808 by Piotto \ea (2007), who
resolve three main sequences in the cluster for a single turnoff (see
figure~\ref{fig:NGC2808}).  This remarkable observation is consistent
with multiple stellar populations of approximately the same age with
varying helium abundances (see inset panel), as outlined initially by
D'Antona \ea (2005).  Such multiple sequences, including those along
the subgiant branch which may imply age variations, have now been
observed in the H--R diagrams of several massive globular clusters
such as NGC~1851, NGC~6388, NGC~6656, NGC~6715 and 47~Tuc (Milone \ea
2008; Anderson \ea 2009; Piotto 2009).  For the most massive Milky Way
globular cluster, $\omega$~Cen, Bedin \ea (2004) demonstrated multiple
turnoffs, subgiant branches and a bifurcation in the main sequence
(see also Piotto \ea 2005; Villanova \ea 2007).  At younger ages,
several LMC clusters have been shown to contain two distinct turnoff
branches separated by an age difference of 300~Myr (Mackey \ea 2008).
When combined with internal spreads in metal abundance, these new
findings are inconsistent with the picture of globular cluster
formation and evolution that has been held for so long.  An accurate
understanding of stellar evolution within these systems will require a
better understanding of mass-loss prescriptions (see
\S\,\ref{massloss}), new chemical-enrichment mechanisms and possibly
modelling with multiple star-formation epochs.

\section{Pushing H--R diagrams to the limit: the complete stellar inventory of 
a star cluster} \label{inventory}

Even the most accurate photometric studies of the nearest star
clusters have been unable to catalogue stars along the entire main
sequence, and rarely probe to the photometric depths required to
detect even the brightest of the remnants of stellar evolution for the
bulk of all stars, white dwarfs.  At very low masses, the
mass--luminosity relation for hydrogen-burning stars is very steep,
and therefore a small change in stellar mass yields a large change in
luminosity.  At solar metallicity, a 0.1~M$_\odot$ star has $M_{V} =
15.9$ mag (Chabrier \ea 2000), giving it a faint optical apparent
magnitude of $V > 27$ mag at a distance of just 2~kpc.  Compounding
the faintness of these stars, the veil of foreground and background
objects also increases with photometric depth, making it difficult to
isolate a clean lower main sequence in most studies.  For the first
time, these obstacles have been crossed, leading to the most accurate
and complete H--R diagram ever published for a star cluster.  This, a
126-orbit {\sl HST}/ACS integration of the globular cluster NGC~6397
(Richer \ea 2006, 2008), has led to photometry of the lowest-mass
hydrogen-burning stars that exist in such systems.  We describe this
remarkable data set here, as it represents a backdrop for several of
the studies in the subsequent sections.

\begin{figure}[ht]
\begin{center}
\leavevmode 
\includegraphics[width=11.0cm]{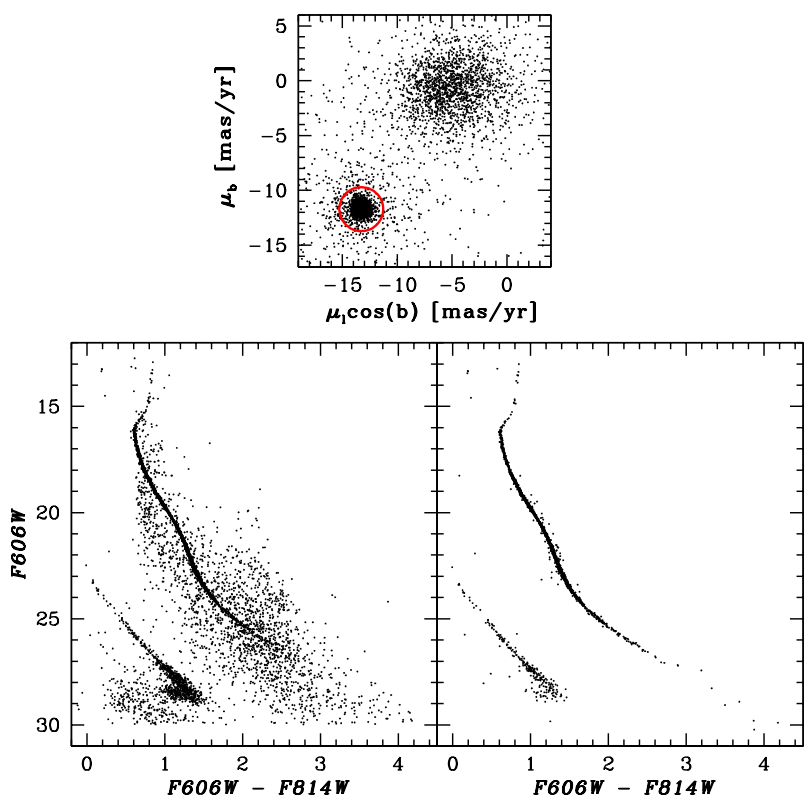}
\end{center}
\caption{The proper-motion {\it (top)} and H--R diagram {\it (bottom)}
of the nearby globular cluster NGC~6397 from a 126-orbit exposure with
{\sl HST}/ACS (Richer \ea 2008).  On the main sequence, the
proper-motion-cleaned data set {\it (right)} extends to the
hydrogen-burning limit at $m_\mathrm{F606W} = 30$ mag (these are the
stars within the red circle in the proper-motion diagram).  The
complete white-dwarf cooling sequence is also characterized at the
faint blue end of H--R diagram, extending almost six magnitudes down
to $m_\mathrm{F606W} = 29$ mag {\it (left)}.  These results are
discussed in \S\,\ref{inventory}.
\label{fig:N6397lowerMS}}
\end{figure}

We present the H--R diagram of NGC~6397 in
figure~\ref{fig:N6397lowerMS}.  In the left panel, a tight sequence
representing the cluster is clearly delineated from the general field
contamination down to $m_\mathrm{F606W} = 26$ mag.  At this point on
the main sequence, the Dotter \ea (2008) mass--luminosity relation
for the metallicity of NGC~6397, [Fe/H] $= -2.03 \pm 0.05$ dex
(Gratton \ea 2003), indicates a mass of 0.092 M$_\odot$, 10\%
higher than the expected hydrogen-burning limit at 0.083~M$_\odot$
(see also King \ea 1998).  Lower-mass stars on the stellar
sequence of NGC~6397 are isolated from the line-of-sight foreground
and background stars by proper-motion selection.  In the top panel,
the astrometry of all stars in the field is measured by comparing
these ACS observations to archival WFPC2 data over 60\% of the field
of view (King \ea 1998).  The tight clump in this diagram at
$\mu_{l} \cos(b) = -13.27 \pm 0.04$~mas~yr$^{-1}$ and $\mu_{b} =
-11.71 \pm 0.04$~mas~yr$^{-1}$ (Kalirai \ea 2007{\it b})
represents the NGC~6397 stars, and the corresponding H--R diagram for
just these stars is shown in the right-hand panel.  The stellar main
sequence of the cluster extends another four magnitudes down to
$m_\mathrm{F606W} = 30$ mag.  The data are still 75\% complete at this
limit, below which no stars are detected (see Richer \ea 2006,
2008).  Therefore, this represents a detection of the hydrogen-burning
limit, a fundamental prediction of stellar evolution theory, at a mass
of $M \sim 0.083$ M$_\odot$, based on the Dotter \ea (2008)
models, and at a similar mass based on the Baraffe \ea (1998)
models.

Above the hydrogen-burning limit, the main sequence illustrated in
figure~\ref{fig:N6397lowerMS} extends 14 magnitudes to the present-day
turnoff of the cluster at $m_\mathrm{F606W} = 16$ mag.  In Richer \ea
(2008), these data are compared to two independent stellar evolution
models to yield valuable insight into the parameter space (e.g., mass
ranges) over which the models match the observations versus regions
where there is a mismatch.  Specifically, these data allow tests of
the detailed shape of the main sequence itself, through comparison of
observed slope changes along the main sequence to those predicted by
the models.  For this work, Dotter \ea (2007) recently constructed a
new grid of stellar evolution tracks, isochrones, luminosity functions
and synthetic horizontal-branch models to compare to such H--R
diagrams as this and those in the ACS Globular Cluster Survey (e.g.,
figure~\ref{fig:8cmds2}).  Finally, we note that comparison of stellar
models to these clean sequences also yields an estimate of the stellar
initial mass function in these environments by relating the
present-day observed mass function in specific fields to the global
primordial function through dynamical simulations of the cluster (see
Richer \ea 2008).

At an age of 12~Gyr, the present-day turnoff mass of a globular
cluster such as NGC~6397 is $\sim 0.8$ M$_\odot$.  All stars that were
more massive than this limit have exhausted their core-hydrogen supply
and evolved to post-main-sequence evolutionary phases (e.g., the
subgiant and red-giant phases, which can be seen in
figure~\ref{fig:8cmds2}).  The final state of this evolution for 98\%
of all stars is white dwarfs.  These stars are the direct products of
helium burning and so their core composition is C and O.  The stars
contain no nuclear fuel to sustain fusion reactions.  As a white
dwarf, a star will simply emit light as it cools and becomes dimmer as
time passes.

Star clusters are privileged sites for studying the evolution of
main-sequence stars into white dwarfs (see, e.g., the review by
Moehler \& Bono 2009).  Although all stars formed at the same time,
the most massive objects exhausted hydrogen rapidly and formed white
dwarfs earlier, and thus have now cooled to fainter magnitudes.  Given
the continuous stream of stars evolving off the main sequence, a
white-dwarf cooling sequence will form at the faint blue end of the
H--R diagram with hotter, newly formed white dwarfs at the top of the
sequence.  Just as discussed above for the faintest main-sequence
stars, the {\sl HST}/ACS observations of NGC~6397 shown in
figure~\ref{fig:N6397lowerMS} unveil the complete white-dwarf cooling
sequence of the cluster, stretching almost six magnitudes in the H--R
diagram.\footnote{The faintest white dwarfs are missing from the
proper-motion-cleaned H--R diagram in figure~\ref{fig:N6397lowerMS} as
these stars were not detected in the shallower, earlier-epoch WFPC2
observations.}  The detection of these stars provides a complete
stellar picture of this globular cluster and leads to several
interesting findings.  For example, modelling of the cooling sequence
of the white dwarfs in the cluster yields an extremely accurate age
measurement for this globular cluster, $t = 11.47 \pm 0.47$~Gyr
(Hansen \ea 2007),\footnote{This was recently updated to an age of
$\sim 12.21 \pm 0.35$~Gyr with the inclusion of a new opacity source,
i.e., the red wing of Lyman $\alpha$ in the white-dwarf cooling models
(Kowalski 2007; B.\ Hansen 2009, personal communication)} the
structure in the cooling sequence verifies theoretical predictions of
the colours of very cool white dwarfs (e.g., the blue hook; see
Bergeron \ea 1995; Hansen 1999; Saumon \& Jacobson 1999), and the
connection between the main sequence and the white-dwarf cooling
sequence allows the first probe of the stellar mass function above the
present-day turnoff of a globular cluster (e.g., through the
initial--final-mass relation; see below and Richer \ea 2008).

\section{Post-main-sequence evolution and stellar mass loss} \label{massloss}

Correctly characterizing the amount of mass loss that stars suffer
through post-main-sequence evolution represents one of the most
important and fundamental goals of stellar astrophysics.  Unlike life
on the main sequence, the total post-main-sequence evolutionary
lifetime of most stars up to the tip of the asymptotic giant branch
constitutes only 5\% of the total stellar lifetime, and therefore this
represents a very dynamic stage in a star's life.  One very important
implication of accurately constraining chemical yields from stars
during post-main-sequence evolution has already been discussed,
namely, the pollution of helium from winds of massive asymptotic giant
branches into the intracluster medium and on to unevolved stars.  Even
prior to this evolutionary stage, the mass-loss rate for low-mass
stars on the first-ascent red-giant branch can drastically affect the
eventual fate of an evolving star.  For example, the total integrated
red-giant-branch mass loss can affect the location from which a star
leaves the red-giant branch (e.g., Castellani \& Castellani 1993;
D'Cruz \ea 1996; Kalirai \ea 2007{\it a}) and therefore alters the
upper red-giant-branch luminosity function.  Equally important, higher
rates of mass loss lead to hotter exposed stars, which will occupy a
bluer position on the subsequent core-helium-burning horizontal branch
(Rood 1973).  Extremely high levels of mass loss on the red-giant
branch can also lead to stars with very thin hydrogen envelopes that
bypass both of these phases and evolve directly to the white-dwarf
cooling sequence with helium cores (e.g., Hansen 2005; Kalirai \ea
2007{\it a}), or those that experience late flashes (see, e.g., D'Cruz
\ea 1996).

\begin{figure}
\includegraphics[width=8cm, angle=270, bb=0 -50 600 350]{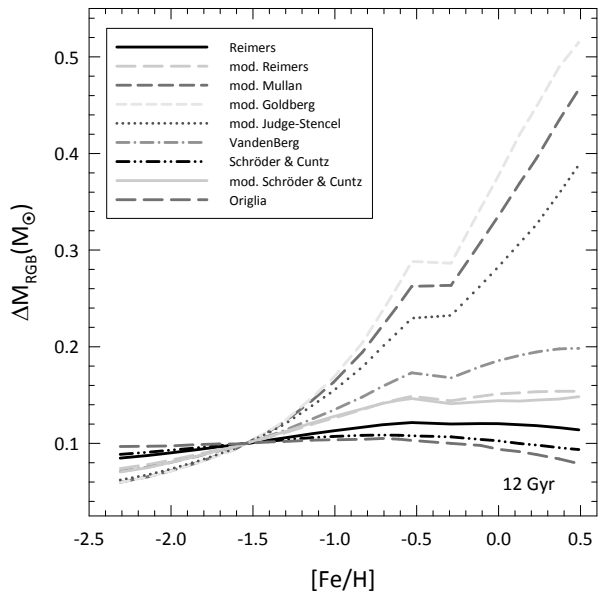}
\caption{Behaviour of several red-giant-branch mass-loss prescriptions
as a function of metallicity for stars with a fixed age of 12~Gyr (see
Catelan 2009 for more information and full references).  The total
mass loss on the branch has been normalized in each model to
0.10~M$_\odot$ at [Fe/H] $= -1.54$ dex to highlight the different {\it
behaviours} of these relations with respect to metallicity.
\label{fig:masslossmetallicity}}
\end{figure}

Unfortunately, theoretical predictions of post-main-sequence stellar
mass loss are difficult to calculate due to insufficient understanding
of mass-loss mechanisms on the red-giant branch and at the helium
flash, and also to the unknown number of thermal pulses on the
asymptotic giant branch (see, e.g., Habing 1996; Weidemann~2000).  The
mass-loss rates depend on the assumed composition of the dust grains,
the dust-to-gas ratio (which likely correlates with metallicity), the
expansion velocity of the stellar envelope and the
temperature/luminosity of the star (see, e.g., Groenewegen 2006).
Given our lack of knowledge of these basic stellar properties,
variations in different mass-loss recipes are large (see
figure~\ref{fig:masslossmetallicity}), leading to large uncertainties
in our basic predictions from stellar evolution models and, therefore,
our interpretation of the properties of stars in these phases today.

The eventual fate of most stars after suffering through mass-loss
stages on the red- and asymptotic giant branches is the white-dwarf
cooling sequence.  Therefore, imprinted within the properties of white
dwarfs, such as their masses, are the integrated events that their
main-sequence progenitors evolved through.  A connection between the
final remnant mass and the initial progenitor mass represents a
powerful probe of mass loss, and is therefore one of the most
important relations in all of stellar astrophysics.  In addition, such
a relation can only be constructed for white dwarfs that are members
of star clusters, as the total stellar lifetime (e.g., the age of the
cluster) is known for each remnant.  Building this fundamental link
requires a step beyond photometric observations.  As most white dwarfs
are degenerate A-type stars, their spectra exhibit the Balmer sequence
of absorption lines, heavily broadened due to the intense pressure on
the surface (i.e., a white dwarf is one million times denser than a
typical rock on Earth).  Measuring these Balmer lines through
spectroscopic observations and reproducing them with atmosphere models
(e.g., Bergeron \ea 1992) yields the masses and cooling ages of the
remnants (i.e., the time since the star left the tip of the asymptotic
giant branch), and the difference between this age and the cluster age
represents the main-sequence lifetime of the progenitor star up to the
tip of the asymptotic giant branch.  Therefore, the initial mass of
the star becomes known through the mass--main-sequence-lifetime
relation for stars of the metallicity of the cluster.

\begin{figure}[ht]
\begin{center}
\leavevmode 
\includegraphics[width=8.0cm]{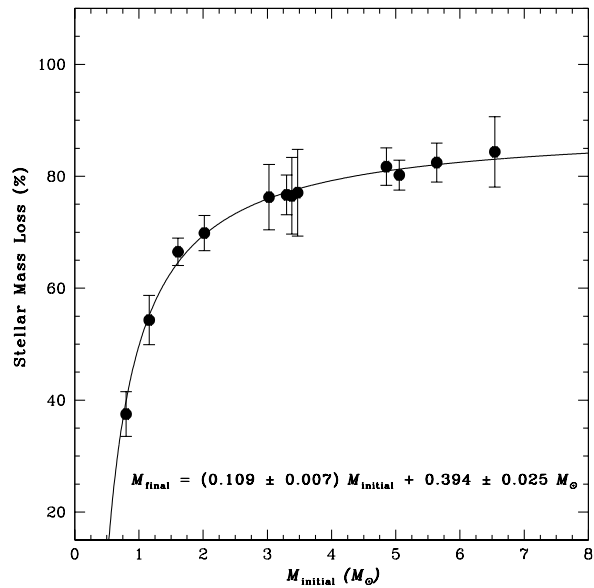}
\end{center}
\caption{Total integrated mass loss through stellar evolution,
constrained from connecting white-dwarf mass measurements in open
clusters to their progenitor masses (see \S\,\ref{massloss}).  The
solid line represents a fit to the data points.
\label{fig:justmassloss}}
\end{figure}

The first efforts to construct this mapping, called the
initial--final-mass relation, date back to Weidemann~(1977).  This was
followed by a two-decade-long effort, primarily by D.\ Reimers and D.\
Koester (see Koester \& Reimers 1981, 1985, 1993, 1996; Reimers \&
Koester 1982, 1989, 1994; Weidemann \& Koester 1983), also including
studies by Weidemann (1987, 1997) and Jeffries (1997).  A review of
the earlier work is provided in Weidemann~(2000) and a compilation of
more recent results is presented in Ferrario \ea (2005), excluding
very recent studies by Dobbie \ea (2006, 2009), Williams \& Bolte
(2007), Williams \ea (2009) and Kalirai \ea (2007{\it a}, 2008,
2009{\it a}).  The synthesis of all of these studies is presented in
figure~\ref{fig:justmassloss}, which displays the total integrated
mass loss as a function of the initial mass of stars (note, most of
the host star clusters of these white dwarfs have solar metallicity).
Clearly, the trend indicates that more massive stars will lose a
higher fraction of their total mass in post-main-sequence evolution,
yet still form more massive white dwarfs.  To quantify stellar
mass-loss rates, the white-dwarf data suggest that, for the most
massive main-sequence stars that will form white dwarfs, the
post-main-sequence stellar yield is about 85\%, and this decreases
smoothly to $\sim 75$\% for intermediate-mass stars with $3 <
M_\mathrm{initial} < 4$ M$_\odot$.  A more rapid decline is seen for
stars with $M \lesssim 2$ M$_\odot$.  At this mass, stars will lose
$\sim 70$\% of their total mass. However, this decreases down to just
$\sim 55$\% for stars approximately the mass of the Sun.

Although the dependence of mass loss on initial mass is well
constrained from figure~\ref{fig:justmassloss}, there are as yet few
constraints on the dependence on metallicity (e.g., to constrain the
theoretical predictions illustrated in
figure~\ref{fig:masslossmetallicity}).  From measuring infrared
excesses from dust in red giants (wind-driven mass loss), Origlia \ea
(2007; 2009, personal communication) find no correlation of increased
mass loss from [Fe/H] $= -2.0$ dex systems to [Fe/H] $= -0.7$ dex
systems.  Kalirai \ea (2009{\it a}) also verify this with a comparison
of the masses of white dwarfs in the metal-poor globular cluster M4 to
a (small) extrapolation of the solar-metallicity initial--final-mass
relation to $M_\mathrm{initial} = 0.8$ M$_\odot$.  Formally, they find
the masses of white dwarfs forming in Population II systems to be $M =
0.53 \pm 0.01$ M$_\odot$. This suggests that the processes of stellar
mass loss in post-main-sequence evolution will drive away one third of
the mass of stars, in excellent agreement with theoretical
predictions, which have long estimated that 0.8~M$_\odot$ stars should
produce $0.51 < M < 0.55$ M$_\odot$ white dwarfs (Renzini \&
Fusi~Pecci 1988; Renzini \ea 1996).\footnote{Moehler \ea (2004)
reported the masses of white dwarfs in NGC~6752 to be 0.53~M$_\odot$,
using a combination of spectroscopy and photometry.}  Interestingly,
for the [Fe/H] $= +0.40$ dex metallicity cluster NGC~6791, Kalirai \ea
(2007{\it a}) measure a mass distribution of white dwarfs peaked at
significantly lower masses than expected, suggesting evidence of
enhanced mass loss at supersolar metallicities.

The characterization of white-dwarf populations in star clusters with
different properties (e.g., metallicity) can clearly influence our
understanding of mass-loss rates and the trends of those rates with
environment.  As we discuss in \S\,\ref{future}, several new imaging
and spectroscopic projects are likely to yield abundant data in this
field in the coming decade.

\section{Stellar death and the dynamical evolution of clusters} \label{binary}

Stars within clusters are born, evolve and then die in some
fashion. How they die turns out to be important for the entire
cluster.  Massive stars may end their lives as black holes and if the
cluster is able to retain these objects, the black hole can accrete
matter, grow and eventually dominate the cluster dynamics in the core.
A more modest star will terminate its existence as a neutron star,
which may make its presence known as an X-ray source.  A collection of
these in the cluster core can also eventually provide an important
source of heating for the cluster.  As we have already discussed,
lower-mass stars terminate their lifes as white dwarfs and other than
the dynamical effect of them losing up to 90\% of their mass (this
mass is likely lost from the cluster) and decreasing the overall
cluster potential, they have not been suspected of playing a major
role in the ongoing cluster dynamics.  Recent observations and
simulations have changed this picture, however, so that it now appears
that white dwarfs are significant players in the dynamical evolution
of star clusters. There are a number of important players in this
story. Therefore, let us see how each of these contribute to the
overall picture.  The cast of characters include the cluster binary
frequency, the perceived need for intermediate-mass black holes
(IMBHs) in some globular clusters and a new scenario which
incorporates the discovery that white dwarfs are apparently given a
kick during their formation.

\subsection{The cluster binary frequency}

There is a general {\it misconception} that the binary frequency in
the disc of the Galaxy is high.  While it is true that it is elevated
($>50$\%) among high-mass disc stars, for low-mass stars (like those
currently on the main sequence in globular clusters) the frequency is
low, i.e., just 30\% of M dwarfs in the disc are in binary systems
(Lada~2006) and the fraction is even lower in the Galactic halo.  This
incorrect belief of a high binary frequency among low-mass field stars
may have been carried across to globular clusters, where it has been
the accepted paradigm for many years that the binary fraction is
20--100\%.

\begin{figure}[ht]
\begin{center}
\leavevmode 
\includegraphics[width=8.0cm]{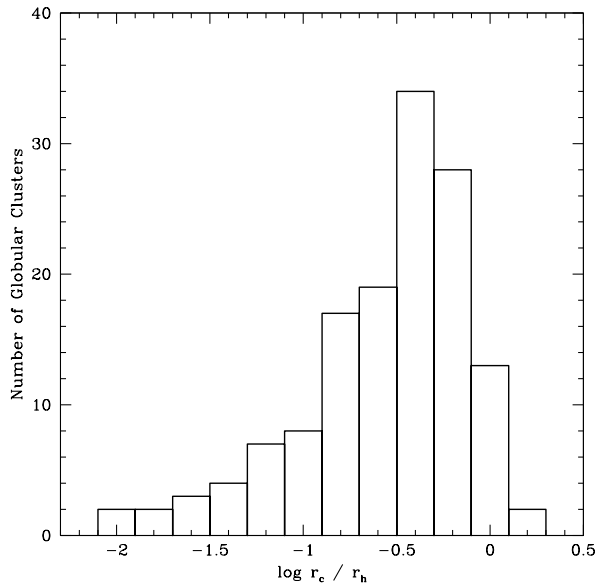}
\end{center}
\caption{Distribution of the ratio of the core to half-mass radii for
globular star clusters.  All data were taken from the Harris (1996)
catalogue.  Note that the majority of clusters have large
$r_\mathrm{c}/r_\mathrm{h}$ ratios, with fully 55\% having a ratio in
excess of 0.3.
\label{fig:corehalfradii}}
\end{figure}

An examination of the distribution of core to half-mass radii
($r_\mathrm{c}/r_\mathrm{h}$) in globular star clusters indicates that
many clusters have large core radii relative to their half-mass radii:
fully 50\% have this ratio $>0.3$ (see
figure~\ref{fig:corehalfradii}).  So, most clusters do not currently
appear to be in a core-collapsed (or binary-burning) phase of
evolution. They may still be in the standard initial contraction phase
or they may have had their core sizes enhanced through some excess
energy source in the cluster.  $N$-body or Monte Carlo models (e.g.,
Heggie \ea 2006) demonstrate that, with zero binaries, a cluster
rapidly goes into core collapse and only with an appreciable
primordial binary fraction ($>10$\%) could core collapse be avoided or
delayed for a time comparable to the age of the Universe.  So, putting
an appreciable binary fraction into the models helps bring the
theoretical structural picture of globular clusters closer to the real
data, although even with 100\% binaries, very large core radii could
not be achieved in dynamical simulations. This was demonstrated early
on, for example by Vesperini \& Chernoff (1994), who studied the
dependence of $r_\mathrm{c}/r_\mathrm{h}$ on binary frequency and
showed that this ratio was always $<0.05 - 0.08$.  So, the structure
of the bulk of the globular star clusters cannot be explained simply
by assuming a large binary fraction: the cluster must either be
dynamically young or have some new energy source.

In a recent paper, Davis \ea (2008{\it a}) show that beyond the
half-mass radius in NGC~6397, for which we have superb {\sl HST}/ACS
data as discussed earlier, the binary frequency is very low, only
1--2\%.  In fact, a literature search yields that virtually all
determinations of binary fractions at large clustercentric radii
suggest a low value.  This result may be unrelated to the primordial
binary frequency, as it is expected that the binaries will sink to the
cluster centre as they are generally more massive objects.  There is
some justification for this as determinations of the binary frequency
in globular cluster cores suggest values at least 3--5 times higher
than at large radii.
   
The situation was recently clarified by Hurley \ea (2007) who
analysed a 100\,000-object (single stars and binaries) $N$-body
simulation of a globular cluster.  They show convincingly that the
primordial cluster binary frequency was basically preserved outside
the cluster half-mass radius and that new binaries were produced in
the core, resulting in a higher frequency there.  This is an important
result, as it strongly suggests that the primordial frequency is very
low and that binaries cannot play an important role in delaying or
preventing core collapse in most clusters.  Further information on the
importance of the binary fraction of star clusters is provided in
Goodwin (2010).
   
\subsection{Intermediate-mass black holes in globular clusters?}  

With binaries clearly not the solution to large cluster cores, the
paradigm shifts to IMBHs.  These are black holes in the mass range
from roughly 100 to a few times 10\,000~M$_{\odot}$.  Such objects
could form from a stellar-mass black hole that subsequently accreted
material from stars shredded by its tidal field or from runaway
mergers of massive stars in the early cluster history. An extension of
the supermassive black hole--velocity dispersion relation for galaxy
bulges predicts IMHBs of several thousand solar masses for the
velocity dispersions typically encountered in the cores of massive
globular clusters (a few tens of~km~s$^{-1}$; e.g., Valluri \ea
2005).

IMBHs act in a way similar to binaries in heating a cluster.  For a
binary system, an interaction with a third star causes the binary to
harden and its total energy to become more negative, while the
interloper goes off with higher kinetic energy to conserve energy.  An
IMBH acts in much the same way, capturing a star in a tight orbit
while interactions with other stars eventually harden the orbit and
provide excess kinetic energy to the interloping star.  $N$-body
calculations by numerous authors (e.g., Gill \ea 2008; Trenti \ea
2007) show that large cluster cores could easily be obtained with the
presence of an IMBH with a mass of a few percent of that of the total
cluster.  Some authors have even made statements to the effect that an
IMBH was {\it required} in the cores of those clusters exhibiting very
large core radii.
   
An observational signature of the presence of an IMBH in a star
cluster is an increase in the cluster's velocity dispersion near the
core (e.g., Baumgardt \ea 2005).  Observations to search for such
a signature are very demanding, as the cluster core has a very high
stellar density.  This is particularly difficult if the observations
are radial-velocity measurements, where there may be enormous amounts
of scattered light in the slit of the spectrograph and often multiple
stars on the same slit.  Several heroic attempts have been made to
search for this effect in a few clusters such as M15 (Gerssen \ea
2002; van der Marel \ea 2002), $\omega$~Cen (Noyola \ea 2008)
and G1 in M31 (Gebhardt \ea 2002, 2005).  While none of these
cases was rock-solid, they were certainly suggestive of the presence
of IMBHs in some clusters.

\begin{figure}[ht]
\begin{center}
\leavevmode 
\includegraphics[width=8.0cm]{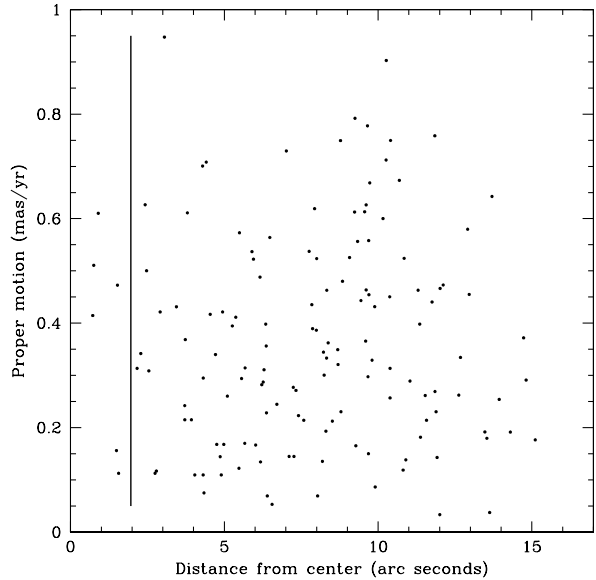}
\end{center}
\caption{Proper-motion distribution of stars within about 15$''$ of
the centre of the globular cluster M71.  These data are from {\sl
Gemini North} ALTAIR/NIRI images with just under a two-year time
baseline between the observations.  The vertical line is the sphere of
influence ($GM_\mathrm{BH}/\sigma^2$) of a 100~M$_{\odot}$ black hole
in the centre of M71 (where $\sigma$ is the stellar velocity
dispersion, $M_\mathrm{BH}$ the black-hole mass and $G$ the
gravitational constant).  Inside this radius, we would expect to see
an increase in the velocities of the stars if there were a black hole
of this mass present.  Clearly, no such upturn is observed.  (These
data were part of the MSc thesis of A.\ Ruberg at the University of
British Columbia.)
\label{fig:pmM71}}
\end{figure}

A few years ago, we began a programme with several students (Andres
Ruberg, Ronald Gagne and Saul Davis) to explore proper motions in the
cores of globular clusters, mainly as a tool to determine a geometric
distance to these systems but also to search for the signature of an
IMBH.  This is a more efficient approach than radial-velocity
measurements as the stellar light does not have to be dispersed
(avoiding overlap as much as possible), observation time is much less
than for spectroscopic measurements and high-spatial-resolution
instruments such as {\sl HST} or adaptive-optics imagers on large
ground-based telescopes could be used to image the cluster cores.  As
a trial case, we used the {\sl Gemini North} adaptive-optics
near-infrared system ALTAIR/NIRI to image the northern metal-rich
globular cluster M71, which has a relatively fluffy core,
$r_\mathrm{c}/r_\mathrm{h} = 0.38$.  The observations were of very
high quality and we resolved the proper-motion {\it dispersion} in the
cluster core ($250 \pm 20 \mu$arcsec yr$^{-1}$).  As we demonstrate in
figure~\ref{fig:pmM71}, the proper motions of stars as a function of
their distance from the centre of the cluster (a new determination of
the cluster centre was made with recent {\sl HST} data) is flat with
no hint of a rise towards the centre (indicating that an IMBH is not
present).  In a recent superb study of $\omega$~Cen, Anderson \& van
der Marel (2009) find a similar result using proper motions from {\sl
HST} imaging and claim that their observations do not support the
conclusions of Noyola \ea regarding the presence of an IMBH in this
cluster.  These two detailed examples certainly weaken the case for
IMBHs in globular clusters.

There are other potential heating mechanisms in globular star clusters
which may be operating and are generally not included in the current
generation of models.  Among these are the presence of stellar-mass
black-hole binaries (Hurley~2007), evaporation of the stellar-mass
black-hole population (Mackey \ea 2007), stellar collisions
(Chatterjee \ea 2009) and core oscillations (Giersz \& Heggie 2009;
Heggie \& Giersz 2009).  However, the observations discussed in the
next section suggest a mechanism that selectively puffs up the young
white-dwarf distribution (Davis \ea 2008{\it b}), implying something
specific happening to them when they are born.

\subsection{A potential new paradigm: kicked white dwarfs}

It seems that we require a new paradigm to explain the core dynamics
in globular star clusters.  Fortunately, one has come forward based on
detailed studies of white dwarfs in a few clusters.  Davis \ea
(2008{\it b}) showed that in our single field in NGC~6397 at about
$2~r_\mathrm{h}$, the young white dwarfs have a more extended radial
distribution than older white dwarfs.  `Young' here means younger than
about three relaxation times (at the location of the field in the
cluster, the relaxation time is $\sim 0.3$~Gyr) while `old' is between
5 and 12 relaxation times. A similar result is also seen in M4
(Davis~2008{\it c}).  The difference in mass between these two
populations is negligible, so the result cannot be due to mass
segregation.  In any case, the effect actually goes the wrong way for
mass segregation. Just before the young white dwarfs were born, they
were the most massive visible stars in the cluster, at about
0.8~M$_{\odot}$, and hence should be centrally concentrated.  The fact
that they were not implies that something happened to these stars when
they were born to fluff up their distribution.  Our interpretation is
that they were given a small natal kick (of about 3--5~km~s$^{-1}$)
and they then took several relaxation times to exhibit the radial
distribution appropriate to their mass.  Monte Carlo simulations seem
to back up these ideas (Fregeau \ea 2009).  It should be clear here
that the only direct evidence for the kick at this time is the more
extended distribution of the younger white dwarfs.  A more definitive
result would be a measurement of the proper-motion dispersion of the
two populations (the young white dwarfs are predicted to have a
smaller velocity dispersion if they were recently kicked;
D. Lynden--Bell 2009, personal communication), but the data currently
are not good enough to detect this in these faint stars (the
measurement will be made in {\sl HST} Cycle 17).

How does this explain the large cluster cores?  A kicked white dwarf
will lose energy to its surrounding stars as it is moving too fast for
its mass in equipartition.  In this sense, it supplies heat to the
cluster, puffing up its core (Heyl~2008{\it b}).  Monte Carlo models
were calculated for clusters with and without white-dwarf kicks
(Fregeau \ea 2009) and those where the kick was present typically
had larger cores by about a factor of ten and generally did not reach
core collapse in a Hubble time. The kick will not, however, be
important in all clusters.  In the models calculated thus far (Fregeau
\ea 2009), those that have velocity dispersions approximately
equal to the size of the kick speed can have their cores expanded by
the kicks at late times by about an order of magnitude.  For those
clusters where the velocity dispersion is significantly larger than
the kick, no effect should be observed.  The younger white dwarfs are,
in fact, more centrally concentrated, as expected in the
high-velocity-dispersion cluster $\omega$ Cen (Calamida \ea 2008).
We have extensive {\sl HST}/ACS imaging of 47~Tuc due to arrive in
Cycle~17 and the prediction is that the young white dwarfs in this
high-velocity-dispersion cluster should also be more centrally
concentrated than the older ones, in sharp contrast to what is seen in
NGC~6397 and M4.

A physical mechanism for a white-dwarf kick is not immediately
apparent.  Some sort of asymmetric mass loss in the late
asymptotic-giant-branch or planetary-nebula phases seems most obvious,
but a kick at the core-helium flash, if it is off centre, may also be
possible (Ivanova~2009, personal communication).  To provide a white
dwarf with a kick of 4~km~s$^{-1}$ would necessitate an asymmetry in
the total amount of mass lost, with typical wind velocities in all
post-main-sequence phases of evolution of about 10\%.  We are
currently investigating this observationally by exploring the radial
distributions of stars in all evolutionary phases beyond the turnoff,
but a theoretical effort here could prove to be quite rewarding.

This kick paradigm is not without its own difficulties.  In a cluster
where the kicks are likely to be important (velocity dispersion $\sim
5$~km~s$^{-1}$, the approximate size of the expected kick), the
crossing time in the cluster will be $\sim 1$ pc / 5~km~s$^{-1}$ $\sim
2 \times 10^5$ years.  If the kick is truly an impulsive event, it
should occur on this or shorter timescales.  Only the core flash, the
second ascent of the asymptotic giant branch or the planetary-nebula
phase are this brief, but most of the stellar mass loss after the
main-sequence turnoff occurs on the red-giant branch, which lasts
about $10^8$ years.  Hansen (2009) also pointed out that wide binaries
containing a white dwarf should have been disrupted by almost any size
kick, yet some wide common-proper-motion pairs containing a white
dwarf do exist.  What is not clear is whether these are true binaries
or the remnants of a disrupted cluster, but further investigation of
this point is important.
      
\subsubsection{Confirmation and future directions of white-dwarf-kick research}

There are several tests of the kick hypothesis that can be made in
addition to the aforementioned tests of the radial-density and
proper-motion distribution of young versus old white dwarfs in
clusters with high velocity dispersions.  In clusters with very low
velocity dispersions, one expects to observe a dearth of white dwarfs
as a kick of 3--5~km~s$^{-1}$ can exceed the escape velocity from the
system.  There are already some claims in the literature that open
clusters possess too few white dwarfs (Weidemann~1977; Williams~2002;
Fellhauer \ea 2003), but caution should be exercised here, as it is
not difficult to hide white dwarfs in binary systems in open clusters
(Hurley \& Shara 2003).  There are a number of globular clusters that
possess very low velocity dispersions (e.g., $\sigma < 1$~km~s$^{-1}$:
Pal~5, Pal~13 and Pal~14). However, most are too distant for
white-dwarf studies.  The nearest systems with low dispersions in
which this test is possible are NGC~6366 ($\sigma \sim
1.3$~km~s$^{-1}$), NGC~5053 ($\sigma \sim 1.4$~km~s$^{-1}$) and
NGC~5466 ($\sigma \sim 1.7$~km~s$^{-1}$), the nearest of which will
have the tip of the white-dwarf cooling sequence at $V \sim 26$ mag.

Another test of the kick hypothesis is to look for radial anisotropy
in the directions of the proper-motion vectors of the young and old
white dwarfs (Heyl 2008{\it a}).  If the young white dwarfs have
recently been kicked, we would expect them to be on more radial orbits
than their more relaxed elders.  Unfortunately, this test was not
possible with our current NGC~6397 data, as we lack a second epoch of
this field with ACS (WFPC2 was used for the proper-motion work).  This
will be remedied in Cycle 17 when we reobserve this field with ACS,
potentially yielding a clean result.

In the end, the effect of a white-dwarf kick is to convert nuclear
energy from the star into dynamical energy for the cluster.  This
illustrates the extreme synergy between stellar and dynamical
evolution and that to obtain a complete picture of the evolution of a
star cluster, one must consider them in detail together.

\section{Conclusions and future outlook} \label{future}

The discussion in this paper focuses on the knowledge we have recently
gained on the evolution of stars, primarily based on deep imaging and
faint spectroscopic studies of star clusters in our Galaxy.  Among the
topics we have discussed are multiple stellar sequences in clusters,
extremely low-mass hydrogen-burning stars, post-main-sequence
evolution and stellar mass loss, white dwarfs and the interplay of
stellar and dynamical evolution within clusters.  As an extension of
these primary studies, it is important to keep in mind a key
implication of these findings as they relate to general astrophysics.

As early as the first resolved imaging studies of Milky Way clusters,
it was understood that these systems are very important tracers of
Milky Way structure and formation processes.  For example, Harlow
Shapley used the distribution of globular clusters on the sky to infer
the location of the centre of the Galaxy (Shapley 1918).  Today, the
comparison of H--R diagrams of Milky Way open and globular clusters to
stellar evolution models produces the calibration that commonly feeds
population synthesis models, which, in turn, are used to interpret the
properties of distant galaxies (e.g., Bruzual \& Charlot 2003).  At
high redshifts, the light that we measure from these unresolved
sources comes from the distribution of stars along post-main-sequence
evolutionary phases, such as the red-giant branch, horizontal branch
and asymptotic giant branch.  Constraining stellar evolutionary
processes, such as mass loss along the red-giant branch, therefore
directly impacts measurements of star-formation histories,
metallicities and mass-to-light ratios of these galaxies.  As we
continue to refine stellar evolution theory with resolved studies of
nearby systems, we must re-define our knowledge of galactic astronomy.

Even the most ambitious projects discussed above, such as the ACS
Survey of Galactic Globular Clusters, represent a pencil-beam study
with respect to surveys that are on the horizon for 2010--2019.  For
example, {\sl Pan-STARRS}, {\sl SkyMapper} and {\sl LSST} will provide
multi-epoch, multifilter, deep homogeneous resolved photometry of most
star clusters in our Galaxy.  It is especially important that these
surveys, unlike the Sloan Digital Sky Survey ({\sc sdss}), target
regions of the Galactic disc to systematically sample open star
clusters.  In this regard, the first wide-field imaging surveys of the
southern hemisphere will image unexplored clusters that surely will
fill empty regions of parameter space (e.g., age and metallicity).
Kalirai \ea (2009{\it b}) present a summary of the type of science
that these surveys may enable as related to star clusters.

A survey such as {\sl LSST} will provide an unimaginable wealth of
observational data to test stellar evolution models.  With a detection
limit of 24--25th magnitude in multiple optical bandpasses in a {\it
single} visit, and a coadded $5\sigma$ depth in the $r$ band of 27.8
mag, {\sl LSST} will yield accurate turnoff photometry for all star
clusters in its survey area out to beyond the edge of the Galaxy.  For
a 12~Gyr globular cluster, this photometry will extend to over three
magnitudes below the main-sequence turnoff at this distance.  For
low-mass stars, previous surveys such as {\sc sdss} and {\sc 2mass}
have yielded accurate photometry of faint M dwarfs out to distances of
$\sim 2$~kpc.  {\sl LSST} will enable the first detection of such
stars to beyond 10~kpc.  At this distance, the colour--magnitude
relation of hundreds of star clusters will be established and permit a
detailed and systematic investigation of variations in the relation
with age and metallicity.  The present-day mass functions of the
youngest clusters will be dynamically unevolved and, therefore,
provide for new tests of the variation in the initial mass function as
a function of environment, down to very-low-mass stars.

Pushing to the hydrogen-burning limit will also greatly benefit from
new infrared observations, for example as proposed in missions such as
{\sl SASIR} (Bloom \ea 2009) and {\sl NIRSS} (Stern \ea 2009).  These
projects aim to image the entire sky in the $JHK$ (and possibly $L$)
bands down to 24th magnitude.  At an age of 1~Gyr, a $M = 0.08$
M$_\odot$ star has $M_V = 19$ mag (Baraffe \ea 1998) and will
therefore be seen by a survey such as {\sl LSST} to 500~pc.  However,
this star has a $V-H$ colour of 8, and is therefore much brighter in
the near-infrared.  With {\sl SASIR} or {\sl NIRSS}, the complete mass
function of all star clusters, down to the hydrogen-burning limit, can
be characterized out to 2.5~kpc.  With the Wide Field Camera 3 (WFC3)
imager on {\sl HST}, and the {\sl James Webb Space Telescope}, such
stars can be easily probed in star clusters at distances of several
tens of kpc.  These observations will also extend beyond the stellar
and substellar threshold and characterize L and T dwarfs. This will
make feasible a new age indicator for star clusters: the magnitude
difference between the onset of brown dwarfs and the hydrogen-burning
limit.

Similar to the wealth of new data that is expected for low-mass
hydrogen-burning stars from these surveys, future observations of
white dwarfs will allow unprecedented studies.  For the field
population, a total of 20\,000 white dwarfs are currently known (e.g.,
{\sc sdss}; Eisenstein \ea 2006).  {\sl LSST} alone will increase the
total sample size of white dwarfs in the Milky Way to over 50~million
(Kalirai \ea 2009{\it c}).  The bright tip of the white-dwarf cooling
sequence is located at $M_V \sim 11$ mag and will be seen in clusters
out to 20~kpc.  For a 1~Gyr cluster, the faintest white dwarfs have
cooled to $M_V = 13$ mag and will be detected in clusters out to
8~kpc.  These white-dwarf cooling sequences not only provide direct
age measurements (e.g., Hansen \ea 2007) for the clusters (and
therefore fix the primary parameter in the theoretical isochrone
fitting, allowing secondary effects to be measured), but can also be
followed up with current ({\sl Keck, Gemini, VLT} and {\sl Subaru})
and future (e.g., {\sl TMT, GMT} and/or {\sl E--ELT}) multi-object
spectroscopic instruments to yield the connection between initial and
final masses over a wide range of environments.  Finally, the synoptic
nature of several future missions will permit proper-motion separation
of cluster sequences from field-star contamination, permitting cleaner
studies of stellar evolution in large samples of these systems.

\begin{acknowledgements}
We consulted a large number of people while writing this paper.  We
especially wish to thank A.\ Dotter, M.\ Catelan, B.\ Hansen, J.\ Heyl
and A.\ Sarajedini for extensive discussions related to the themes
discussed in this paper.  JSK's research is supported in part by a
grant from the Space Telescope Science Institute's Director's
Discretionary Research Fund. The research of HBR is supported by
grants from the Natural Sciences and Engineering Research Council of
Canada. He also thanks the Peter Wall Institute for Advanced Studies
at the University of British Columbia for the award of a Distinguished
Professorship, which allowed him the time to write this review.
\end{acknowledgements}

\end{document}